\documentstyle[osa,manuscript,psfig]{revtex}

\begin{document}
\title{Longitudinal polarization of hyperons in high $p_\perp$ 
jets in singly polarized $pp$ collisions at high energies}
\author{Xu Qing-hua, Liu Chun-xiu\footnote{Present address:
 Institute of High Energy Physics, Beijing 100039, China}
 and Liang Zuo-tang\\
Department of Physics, Shandong University,
Jinan, Shandong 250100, China}
\maketitle

\begin{abstract}
We calculate the longitudinal polarizations of 
hyperons in high $p_\perp$ jets in $pp$ collisions 
in which one of the protons is longitudinally
polarized at RHIC energies 
using different models for the spin transfer in  
fragmentation process.
The results show that the measurements of 
these polarizations can be used to study 
the spin transfer in high energy fragmentation processes 
in general and to test the different models in particular.  
Our results show especially that 
the magnitude of the polarization
of $\Lambda$ is rather small whereas 
that of $\Sigma^+$ is considerably larger 
in the large rapidity region.
The differences between the results from different 
pictures for $\Sigma^+$ polarizations is also 
much larger. 
Hence, the measurement of $\Sigma^+$ polarization 
should be more effective to distinguish between different
models especially the SU(6) or the DIS picture 
for spin transfer in fragmentation processes.
 
\end{abstract}

%\twocolumn\narrowtext
\newpage

\section{Introduction}

Spin effects in fragmentation processes play an important 
role in understanding the spin physics in 
strong interaction. 
With the developments in experiments,  
much attention has been attracted to the studies 
of the different aspects in this connection 
(see, e.g. Ref.[1-\ref{Ma01pp}] and references therein).
Data are now available from different 
lepton-induced reaction experiments [1-\ref{E665}] 
and many of the theoretical studies [\ref{JAF}-\ref{XBJ}] also 
concentrate on those reactions.   
It should be emphasized that the spin effects in the fragmentation 
processes can also be studied in polarized hadron-hadron collisions 
by measuring the properties of the hadrons in high 
transverse momentum $p_\perp$ jets created in these 
reactions [\ref{Florian98}-\ref{Ma01pp}]. 
Here, because of the validity of factorization theorem, 
the hadrons can be regarded as the pure results from the 
fragmentation of the scattered quark whose polarization 
can be determined using perturbation theory in 
quantum chromodynamics (QCD) and the quark distribution 
functions which determine the polarization before the hard 
scattering.  

Compared with those in the lepton-induced reactions, 
the study of hadrons in high $p_\perp$ jets 
in polarized hadron-hadron collisions 
has the following properties. 
(1) Since the involved hard scatterings 
are strongly interacting processes rather than 
the electroweak processes involved in the lepton-induced reactions, 
the corresponding cross section should in general be larger.
Furthermore, the luminosity of the incoming proton beams 
can in general be made higher than that for leptons 
in particular much higher than that for neutrino. 
Hence, the statistics can be improved in particular 
compared with the lepton-induced reactions with neutrino beams.
(2) There are many different hard subprocesses 
which contribute to the production of 
the high $p_\perp$ jets. 
In particular, we have a large contribution from 
gluon interactions. 
This, on the one hand, makes the study more complicated 
than those in lepton-induced reactions, on the other, 
makes it more interesting since we can use it to study 
not only quark but also gluon polarization 
and fragmentation which are even poorly studied yet. 
Hence, it is also very important to know 
whether there exist kinematic regions 
where gluon fragmentation is important 
or where quark fragmentation dominates.
In this case, we can study them separately.

In this paper, we first make an analysis 
of the contributions of the different hard subprocesses 
to the production of different hyperons 
in different kinematic regions
in $pp$ collisions using a Monte Carlo event generator.
Our results show that, 
there do exist kinematic regions  
where contributions from quark fragmentation dominate, 
while in the other gluon fragmentation plays an important role. 
We then calculate the longitudinal polarization of the hyperons 
in the regions where quark fragmentation dominates using the 
same method as we used in studying quark fragmentation in 
the lepton-induced reactions. 
In section 2, we present the results for the Monte Carlo
analysis of the contributions of different hard processes.  
In section 3, we briefly summarize the calculations and the main 
results for the polarizations of the scattered partons 
from different hard subprocesses. 
In section 4, we present the results for the polarizations
of different hyperons from high $p_\perp$ jets. 
A brief summary of the results is given in section 5.

\section{Contributions of different hard subprocesses to hyperon production 
in high $p_\perp$ jets}

We consider the inclusive production of hyperons
with high transverse momentum $p_\perp$ 
in $pp$ collisions at high energy.
Here, the factorization theorem can be applied, hence
the cross section can be written as\cite{Factorm},
\begin{equation}
E\frac{d\sigma}{d^3p}(pp\to HX)=
\sum_{abcd}\int dx_a dx_b f_a(x_a,\mu^2)f_b(x_b,\mu^2)
\frac{1}{\pi z_c}
\frac {d \hat {\sigma}} {d\hat t}(ab \to cd)
D_c^H(z_c,\mu^2).
\label{sigma}
\end{equation}
Here, $f_a(x_a,\mu^2)$ and $f_b(x_b,\mu^2)$ are the unpolarized 
distribution functions of partons $a$ and $b$ in proton
at the scale $\mu$,
$x_a$ and $x_b$ are the corresponding momentum fractions carried
by $a$ and $b$;
$D_c^H(z_c,\mu^2)$ is the fragmentation function of parton $c$
into hadron $H$,
$z_c$ is the momentum fraction of parton $c$ carried by the 
produced $H$;
${d\hat{\sigma}}/{d\hat t}$ is  the
cross section at the parton level and $\hat t$ is the
usual Mandelstam variable of the parton scattering.
The cross section of the hard subprocess 
can be calculated using perturbative QCD.
The summation in Eq. (\ref{sigma}) 
runs over all possible subprocesses. 
To the leading order in pQCD, 
there are following subprocesses that contribute here, i.e.
the scattering between quark(s) and/or antiquark(s),
$qq \to qq$,
$q\bar q\to q\bar q$, 
$\bar q\bar q \to \bar q \bar q$;
the scattering between gluons or that between gluon and quark, 
$qg \to qg$, 
$\bar qg \to \bar qg$, 
$gg \to gg$;
the annihilation processes,
$q\bar q \to gg$, 
$gg\to q\bar q$.
The scattering matrix elements for these subprocesses 
can easily be calculated and the results can e.g. be 
found in Ref. [\ref{QCD}].
For scattering angles less than $\pi/2$ in the center of 
mass frame of the scattering partons, 
the cross section for $gg\to gg$ is the largest, 
and $gq\to gq$ is the second largest one. 
They are much higher than the $qq$ or $q\bar q$ scattering 
in particular at the scattering near $\pi/2$. 
Since the energies of gluons in protons 
are in general smaller compared with 
those for the quarks and/or antiquarks, 
we expect that $gg$ scattering plays an important role 
for jets with moderately large 
transverse momenta $p_\perp$. 
But, for very high $p_\perp$, $qq$ scattering should dominate.

\begin{figure}[h]
\psfig{file=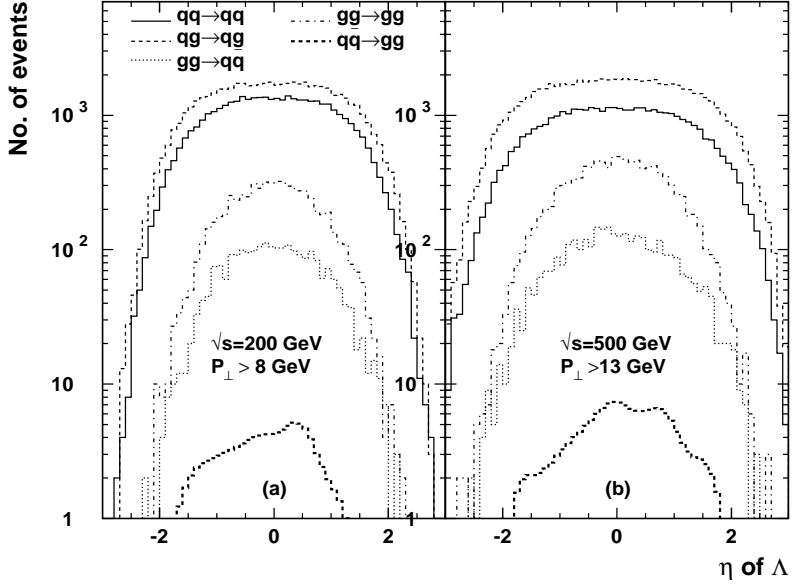,width=12cm}
\caption{Contributions of the different parton level subprocesses
to $\Lambda$ production in $pp \to \Lambda X$.}
\label{jlam}
\end{figure}

\begin{figure}[h]
\psfig{file=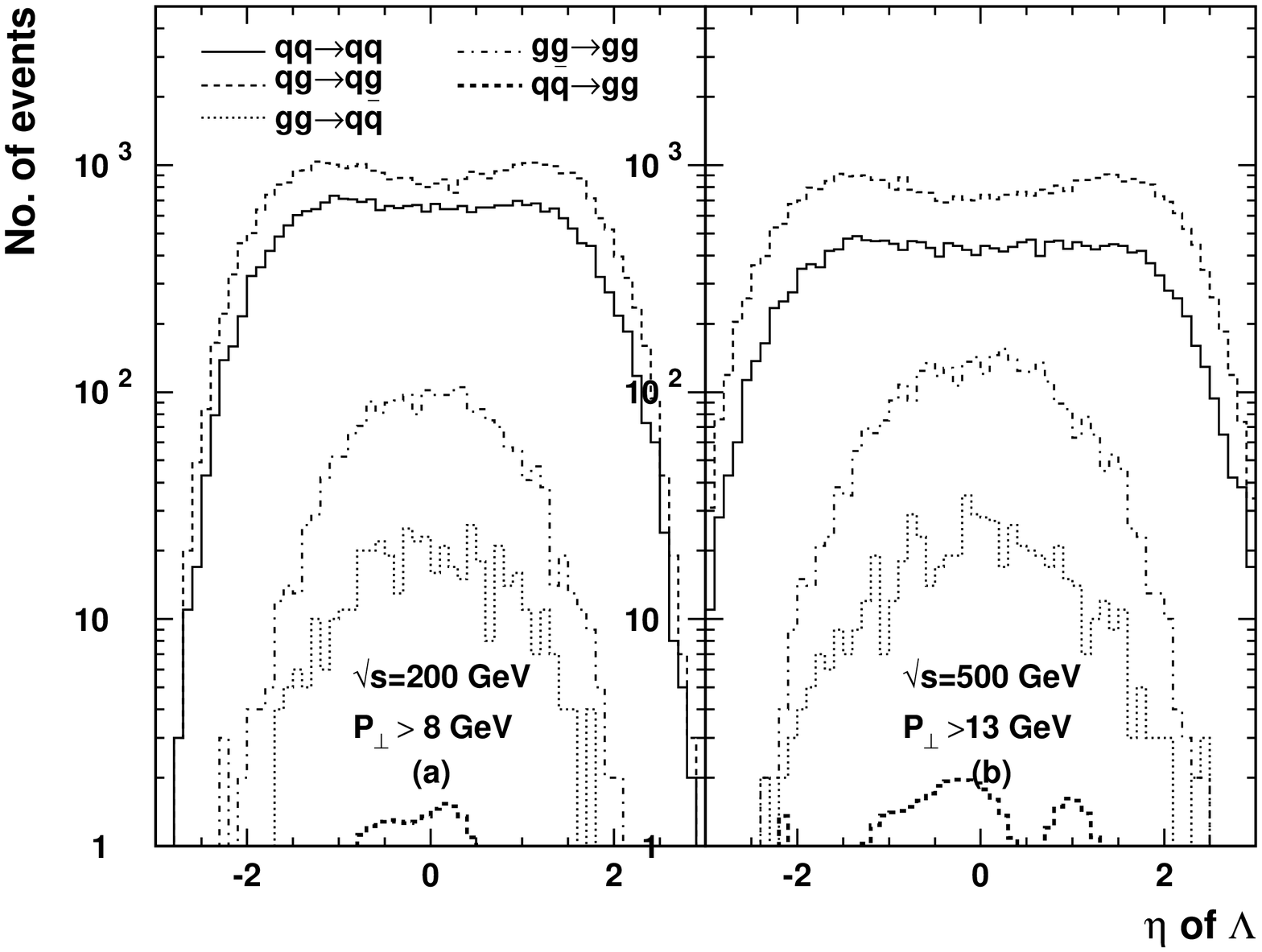,width=12cm}
\caption{Contributions of the different parton level subprocesses
to $\Sigma^+$ production in $pp \to \Sigma^+ X$.}
\label{jsig}
\end{figure}

\begin{figure}[h]
\psfig{file=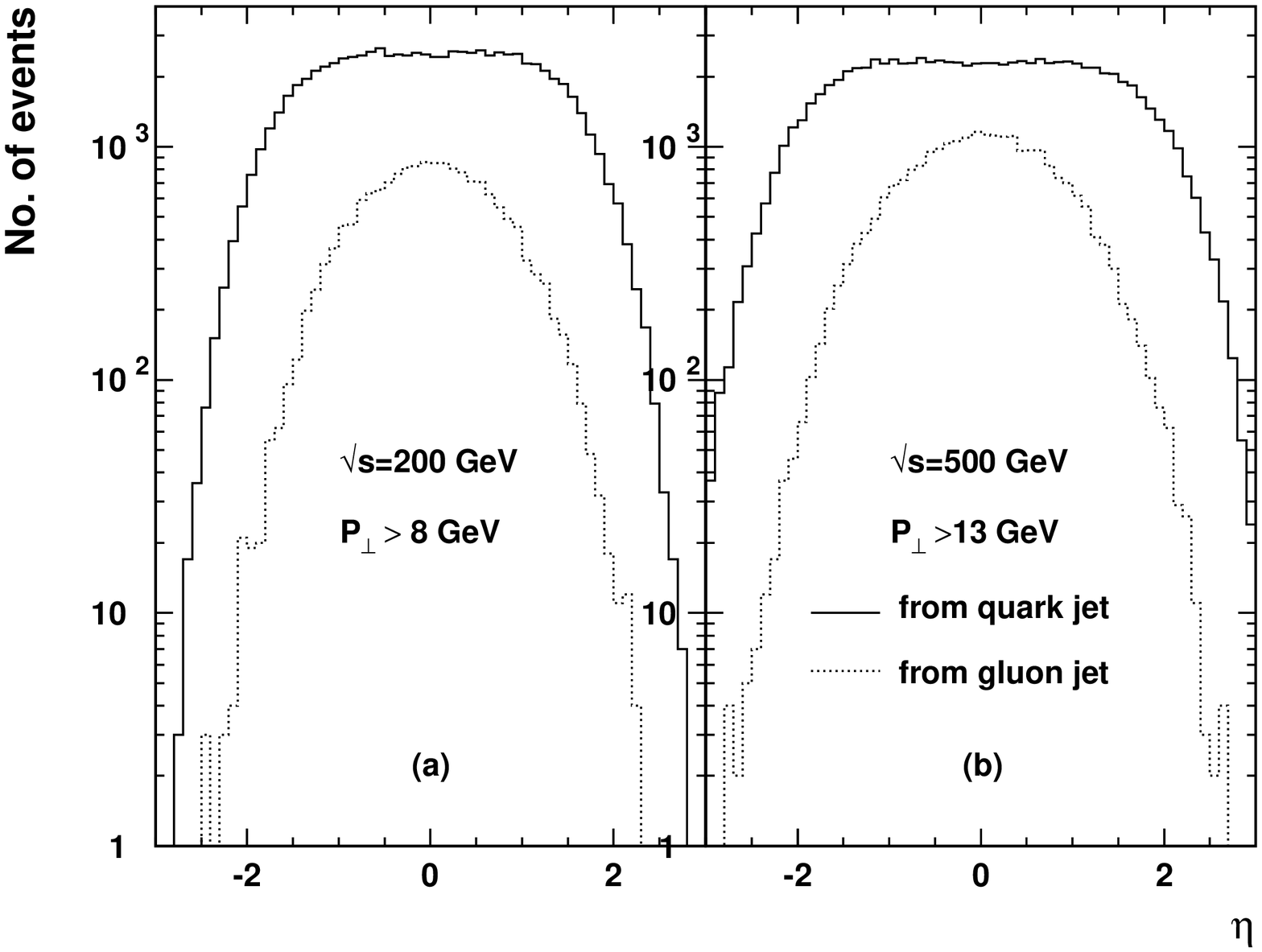,width=12cm}
\caption{The contributions of quark and gluon jets
to $\Lambda$ production in $pp\to\Lambda X$.}
\label{qgjet}
\end{figure}

\begin{figure}[h]
\psfig{file=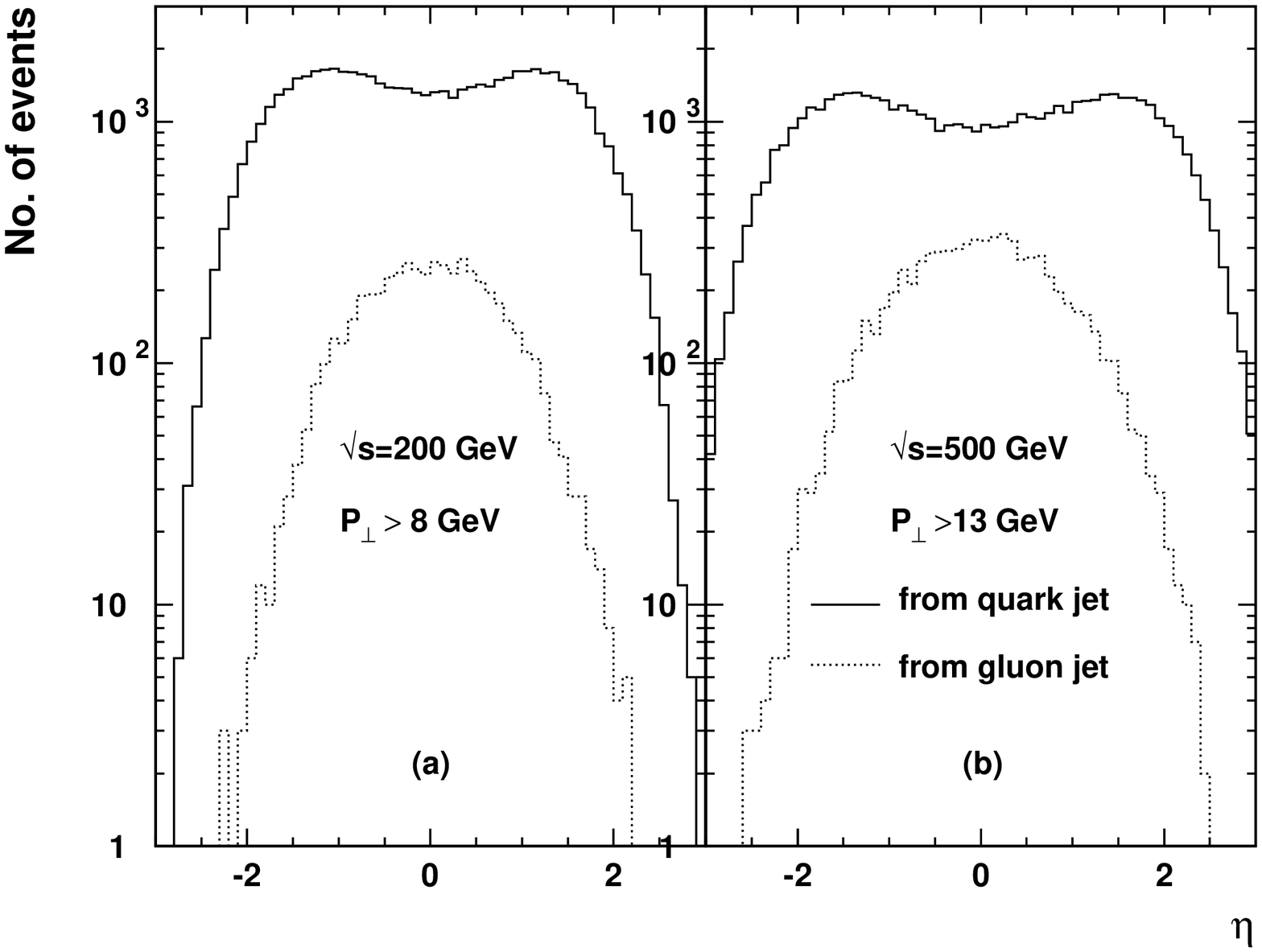,width=12cm}
\caption{The contributions of quark and gluon jets
to $\Sigma^+$ production in $pp\to\Sigma^+ X$.}
\label{qgjetsig}
\end{figure}

\begin{figure}[h]
\psfig{file=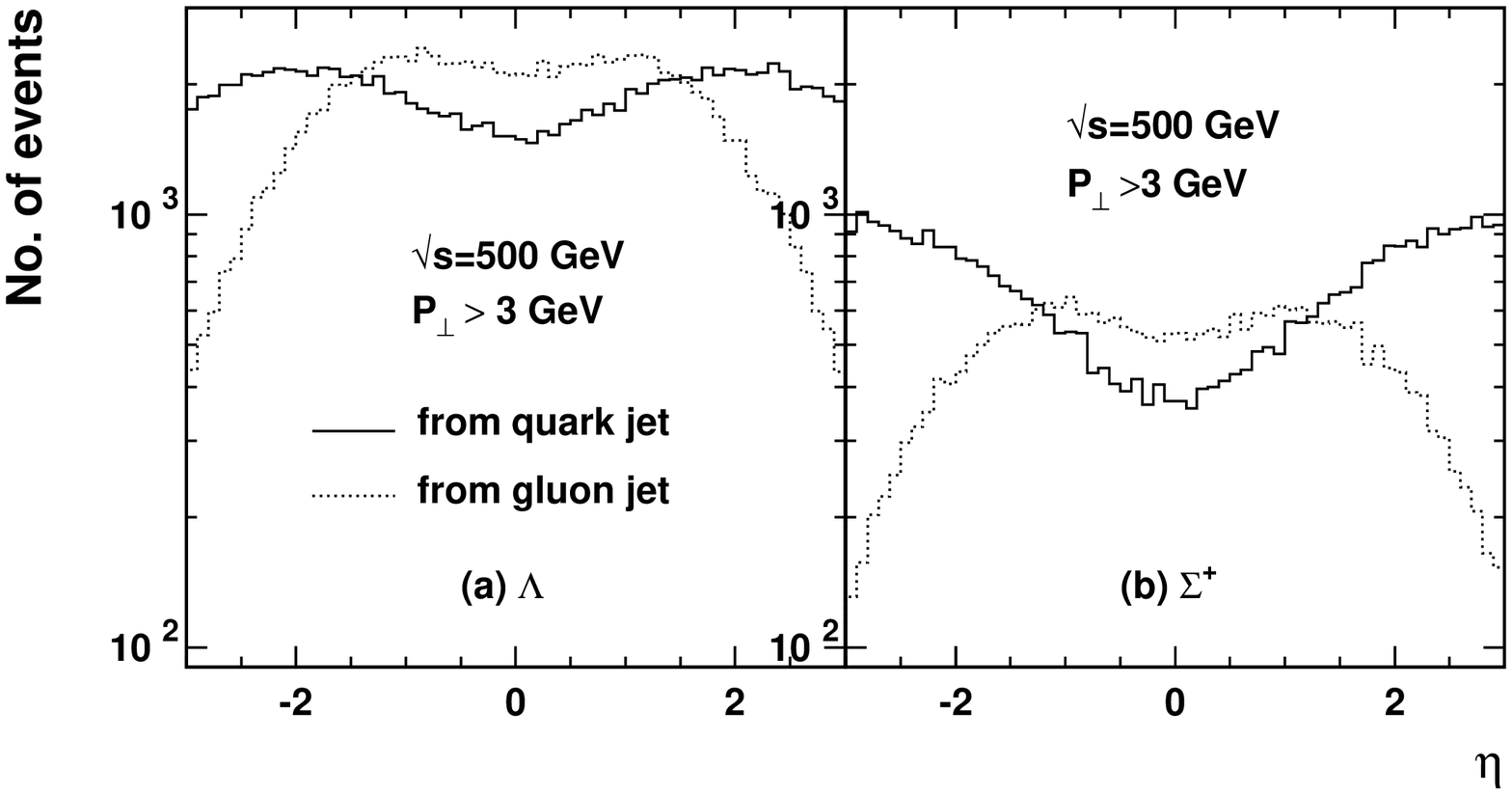,width=12cm}
\caption{The contributions of quark and gluon jets
to $\Lambda$ production (a) and
$\Sigma^+$ production (b) at $p_\perp$$>$3 GeV in $pp\to HX$.}
\label{smpt}
\end{figure}

Taking the fragmentation into account, 
we can estimate the contribution from each subprocess 
to the production of different hyperons. 
Presently, this can be conveniently done using Monte Carlo 
event generators which can give a good fit to the unpolarized data. 
We used {\sc pythia}\cite{lund,PYTHIA} in our calculations and obtained 
the results for $\Lambda$ and $\Sigma^+$ 
as a function of pseudorapidity $\eta$
in Figs.\ref{jlam} and \ref{jsig}.
From these figures, we explicitly see that, 
at high $p_\perp$, e.g.,
$p_\perp$$>$8 GeV at $\sqrt s$=200 GeV
or $p_\perp$$>$13 GeV at $\sqrt s$=500 GeV,
only a few subprocesses dominate.
The contributions of quark scatterings $qq\to qq$ 
(which denotes all the subprocesses 
in which the initial and final partons are all quarks)
and $qg \to qg$ 
are important whereas that from $gg \to gg$ is suppressed significantly.
We also calculated the contributions from quark jets or gluon jets. 
The results are shown in Figs.\ref{qgjet} and \ref{qgjetsig}.
We see that the contribution from gluon fragmentation is 
negligibly small in this kinematic region, 
especially in the large rapidity region (e.g. $\eta$$>$1.5). 
Therefore, the uncertainty of the calculations 
of hyperon polarizations in such kinematic regions 
caused by the unawareness of 
the polarized gluon fragmentation function should
be very small. 
In our calculations which will be presented in Section 4, 
we will just neglect this contribution. 
However, at lower $p_\perp$, the contribution from $gg\to gg$ 
or that from gluon fragmentation is very important.
As an example, we show in Fig. \ref{smpt} the corresponding results
for $p_\perp$$>$3 GeV at $\sqrt s$=500 GeV.
We see that the contribution of gluon fragmentation is higher than 
that of quark in the central rapidity regions. 
Hence, we can study the spin transfer in quark fragmentation 
in the high $p_\perp$ region and that in gluon fragmentation 
in relatively small $p_\perp$ region.

\section{Spin transfer and 
polarizations of the scattered quarks in different 
subprocesses}

Since we are considering only the elementary subprocess
with high $p_\perp$, the scattering matrix elements
can be calculated to high precision using pQCD. 
The spin transfer factor for each subprocess 
can be obtained from the helicity amplitudes.
The results show that, 
for all the different kinds of processes $ab\to cd$ 
(Here, $a$, $b$, $c$ and $d$ denote different partons, 
where $a$ denotes the one from the 
longitudinally polarized incoming proton),
the longitudinal polarization 
can be transferred to the outgoing parton $c$ and $d$. 
The polarization transfer factors are in general different 
for different scattering processes.
But they have a common feature that they are only 
functions of a single variable $y$, 
which is defined as 
$y\equiv k_b\cdot (k_a-k_c)/k_a\cdot k_b$, where 
$k_a$, $k_b$, $k_c$ and $k_d$ are the four-momenta 
of the partons $a$, $b$, $c$ and $d$ respectively.
We summarize these results in Table 1.  

\begin{figure}
\psfig{file=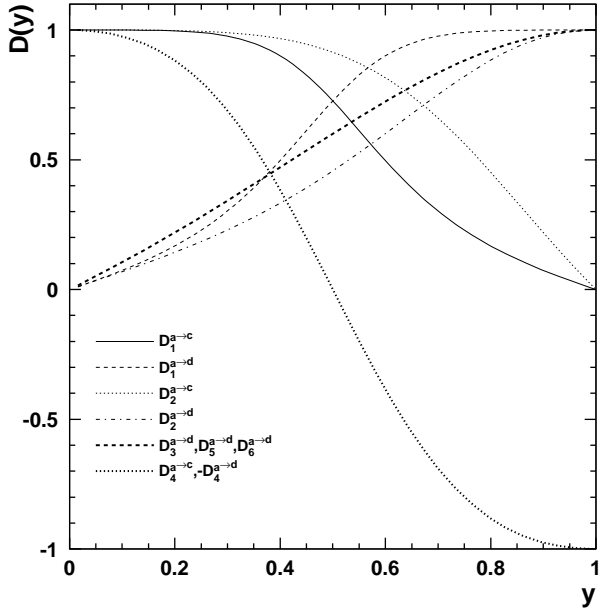,width=10cm}
\caption{Polarization transfer factors
$D^{a\to c}_{(i)}(y)$ and
$D^{a\to d}_{(i)}(y)$ in the $i$-th kind of the
elementary processes $ab \to cd$ in Table 1.}
\label{dy}
\end{figure}

From the table, we see the following:
(i) Since we neglect quark mass, helicity conservation is valid. 
This can be seen from the results 
that $D_{(3)}^{a\to c}=D_{(5)}^{a\to c}=1$, 
and that $D_{(4)}^{a\to c}=-D_{(4)}^{a\to d}$.
(ii) In the scattering processes $qq'\to qq'$ and $qg\to qg$, 
the polarization transfer from one of the colliding objects 
to the other is the same, and they are also the same as 
that in the QED process $e^-\mu^-\to e^-\mu^-$ or 
$e^-q\to e^-q$. 
The result is a monotonously increasing function of $y$, 
which increases from zero to one as $y$ goes zero to one.   
To get a feeling of all these spin transfer factors, 
we show them as functions of $y$ in Fig. \ref{dy}.

Multiplied by the polarizations
of the incoming quarks or gluons, which are 
determined by the helicity distribution functions 
of the partons in proton, 
we obtain the polarizations 
of the outgoing quarks or gluons. 
They can be used to calculate the hyperon polarization 
produced in the fragmentation of these partons.
Clearly, the obtained results depend on the 
polarized parton distribution functions we use.
We will also show how strongly the final 
hyperon polarizations depend on the different sets  
of parameterization \cite{GRSV2000,GRV98,GRSV96,GRV94} 
of these distributions. 
%In the following calculations, we use two different sets of 
%polarized and unpolarized distribution functions.
%In Set I we use the Standard LO set of GRSV2000\cite{GRSV2000}
%for polarized distribution function and GRV98\cite{GRV98} LO set 
%for unpolarized distribution functions; 
%In Set II, Standard LO set of GRSV96\cite{GRSV96} and 
%GRV94\cite{GRV94} LO set are used respectively.  

\section{Longitudinal hyperon polarization 
in $pp \to H X$ in high $p_\perp$ regions}

As we have seen in Sec. 2, for hyperons 
with considerably high transverse momenta, 
quark fragmentation dominates whereas gluon 
fragmentation can be neglected. 
In such kinematic regions, we can calculate 
the hyperon polarization using the same method 
as we used in treating $q_f^0\to H_i+X$ in lepton-induced reactions. 
Measurements of the hyperon polarizations in 
these kinematic regions can be used as further tests of the 
different pictures for spin transfer 
in high energy quark fragmentation processes. 
The experimental studies can 
also be made in other kinematic regions 
where gluon fragmentation plays an important role. 
By comparing the obtained results with those obtained 
in the regions where quark fragmentation dominates, 
we can get useful information for gluon fragmentation.
In this section, we present the calculations and the main 
results in the kinematic regions where quark fragmentation dominates.

\subsection{The calculation method}

The method of calculating the longitudinal polarization $P_{H_i}$
of different hyperon $H_i$ in the fragmentation of a longitudinally
polarized quark $q_f^0$ was outlined in Ref. [\ref{LL2000}]
using the inclusive process
$e^+e^-\rightarrow H_i+X$ as an example. 
It has been applied to $e^+e^-$ annihilations\cite{LL2000} 
and deeply inelastic scattering using different lepton beams\cite{LXL2001}.
We now summarize the main points in the following 
not only for completeness but also for the following reasons: 
Here, we would like to emphasize 
what kinds of inputs that we need in the calculations and  
how we obtain these inputs. 
In this way, we can see from where 
the theoretical uncertainties originate in the calculations 
and how we should reduce these uncertainties by 
doing measurements for different hyperons 
and/or in different kinematic regions.

We consider $q^0_f\to H_i+X$ and divide 
the produced $H_i$'s 
into the following groups:
(a) directly produced 
and contain the $q_f^0$'s; 
(b) decay products of heavier 
hyperons which were polarized before their decays; 
(c) directly produced but 
do not contain the $q_f^0$'s; 
(d) decay products of heavier hyperons 
which were unpolarized before their decays. 
Obviously, hyperons from (a) and (b) 
can be polarized while those from (c) and (d) are not. 
Hence, the polarization of $H_i$ is given by,
\begin{equation}
P_{H_i}={ {\sum\limits_f t^F_{H_i,f} P^{(q)}_f \langle n^a_{H_i,f}\rangle
+\sum\limits_{j} t^D_{H_i, H_j} P_{H_j} \langle n^b_{H_i, H_j}\rangle}
 \over
{\langle n^a_{H_i}\rangle +\langle n^b_{H_i}\rangle +
\langle n^c_{H_i}\rangle +\langle n^d_{H_i}\rangle} }.
\label{polh}
\end{equation}
The different quantities here are defined 
and obtained in the following way: 

(i) $P_f^{(q)}$ is the polarization of $q_f^0$ 
which is determined by the initial conditions 
and the elementary hard scattering processes.

(ii) $\langle n^a_{H_i,f}\rangle$ is the average number of 
$H_i$'s which are directly produced and contain 
$q_f^0$ of flavor $f$, and
$\langle n^b_{H_i,H_j}\rangle$ is that 
from the decay of $H_j$'s which were polarized;
$P_{H_j}$ is the polarization of $H_j$;
$\langle n^a_{H_i}\rangle$,
$\langle n^b_{H_i}\rangle$,
$\langle n^c_{H_i}\rangle$ and $\langle n^d_{H_i}\rangle$
are the average numbers of $H_i$'s 
in group (a), (b), (c) and (d) respectively.
These average numbers of the hyperons of different 
origins are determined 
by the hadronization mechanisms and should be 
independent of the polarization of the initial quarks.
Hence, we can calculate them using a hadronization 
model which give a good description of the unpolarized data. 
We used Lund model\cite{lund} implemented by {\sc jetset} or 
{\sc lepto} or {\sc pythia}\cite{PYTHIA} in our calculations 
for different reactions. 

(iii) $t^F_{H_i,f}$ is the probability for 
the polarization of $q_f^0$ to be transferred 
to $H_i$ in group (a), i.e. to the $H_i$ which contains $q_f^0$, 
and is called the polarization transfer factor, 
where the superscript $F$ stands for fragmentation. 
It is taken as the fraction of 
spin carried by the $f$-flavor-quark 
divided by the average number of 
valence-quark of flavor $f$ in $H_i$. 
This is the place where different pictures come in, 
since the contributions to the hyperon spin from different flavors 
are different in the SU(6) or the DIS picture\cite{BL98,LL2000}.
In the SU(6) picture, these contributions can be obtained 
from the SU(6) wave functions of the hyperons. 
In the DIS picture, for the $J^P=(1/2)^+$ octet hyperons, 
they are obtained from the DIS data on the spin dependent 
structure functions and those on hyperon decay. 
But, for the decuplet hyperons, they are unknown yet. 
In our calculations, we make a rough estimations of these 
contributions by taking them as the same as those in 
the SU(6) picture. 
It will be interesting and important to study 
theoretically which picture 
is applicable in determing $t^F_{H_i,f}$. 
Discussions along this line can be found, for example, 
in [\ref{Ada97}] and the references given there. 

(iv) $t^D_{H_i,H_j}$ is the probability for 
the polarization of $H_j$ to be transferred to 
$H_i$ in the decay process $H_j\to H_i+X$ and 
is called decay polarization transfer factor, 
where the superscript $D$ stands for decay. 
It is determined by the decay process and is independent 
of the process in which $H_j$ is produced.
For the octet hyperon decays, they are 
extracted from the materials in Review of Particle 
Properties\cite{PDG2000}. 
But, for the decuplet hyperons, we have to use 
an estimation based on the SU(6) quark model. 
(The details are given in e.g [\ref{GH93},\ref{LL2000}]).
Hence, we see that decuplet hyperon decay is
a major source of the theoretical uncertainties 
in our calculations of 
the final $P_{H_i}$'s in different reactions.
For them, in the DIS picture, neither $t^F_{H_i,f}$ 
nor $t^D_{H_i,H_j}$ is known yet. 
To reduce this uncertainty, it is important to consider 
the hyperons to which decay contributions are small.

We note also that $t^F_{H_i,f}$ and $t^D_{H_i,H_j}$ 
should be considered as universal in the sense that 
they are the same in different reactions.  
The differences between different processes
come from the polarization $P^{(q)}_f$ of
the fragmenting quark $q^0_f$
and the average numbers $\langle n^a_{H_i,f}\rangle$,
$\langle n^b_{H_i,H_j}\rangle$, $\langle n^c_{H_i}\rangle$,
and $\langle n^d_{H_i}\rangle$ for hyperons $H_i$
from the different origins.
In $pp$ collisions that we are considering in this paper,
$P_f^{(q)}$ is determined by the polarization of the partons before 
the hard scattering and spin transfer factors in the scattering which 
are given and discussed in last section. 

The average numbers $\langle n^a_{H_i,f}\rangle$,
$\langle n^b_{H_i,H_j}\rangle$,
$\langle n^c_{H_i}\rangle$,
and $\langle n^d_{H_i}\rangle$
are determined by the hadronization mechanism
and the relative abundance of different flavors 
that take part in the reactions. 
They should be independent of the process
in which $q_f^0$ is produced and they are also 
independent of the polarization of $q_f^0$.
In $pp$ collisions, 
as we can see from  Eq.(\ref{sigma}), 
these average numbers are determined by 
three factors, i.e., 
the parton distribution functions $f_a(x_a,\mu^2)$ and $f_b(x_b,\mu^2)$,
the cross sections of hard
subprocesses $d\hat{\sigma}/d\hat{t}$, 
and the fragmentation functions $D_c^H(z_c,\mu^2)$.
The cross section $d\hat{\sigma}/d\hat{t}$  
can be calculated with high accuracy from pQCD.
For $f_a(x_a,\mu^2)$, $f_b(x_b,\mu^2)$, and $D_c^H(z_c,\mu^2)$,
although neither of them is theoretically clear yet,
the form of the unpolarized parton densities in nucleon
and that of the unpolarized fragmentation functions
are empirically known to reasonably high accuracy.
We can calculate them using the parameterizations
of the structure functions and the phenomenological
hadronization models.    
Since we are not concerned with the
different correlations among the produced particles,
different hadronization models lead essentially to the same results.
Presently, these results can be obtained
conveniently from the event generators
using Monte Carlo method.
We use the Lund string fragmentation model\cite{lund}
implemented by {\sc pythia}\cite{PYTHIA} in our calculations
for $pp$ collisions.

\subsection{$\Lambda$ polarization in $pp \to \Lambda X$
with longitudinally polarized beam}
 
Most of the studies now available on the spin transfer
concentrate on $\Lambda$ polarization since
$\Lambda$ is most abundantly produced
among all the $J^P=(1/2)^+$ hyperons and
it's polarization can be measured easily through
the decay channel $\Lambda \to p\pi^-$.
Meanwhile, as we have seen already
in $e^+e^-$ annihilation \cite{LL2000} and 
deeply inelastic lepton-nucleon scattering \cite{LXL2001},
the origins of $\Lambda$ are also the most complicated.
They can be not only directly produced from the fragmentation
of $u$, $d$, or $s$ quark, but also from the decays
of many different heavier hyperons,
such as, $\Sigma^0$, $\Xi^{0,-}$,
$\Sigma^*(1385)$, and $\Xi^*(1530)$.
In the large $p_\perp$ regions in $pp$ collisions, 
the fragmentations of the quarks from 
the subprocesses $qg \to qg$ and $qq \to qq$ dominate.
In particular for hyperons with large rapidities, 
the scatterings of quarks with 
large fractional momenta play the most important roles.
Since at large $x$ in proton, 
$u$ quark density is much higher than 
that of $d$, and both of them are much higher than that of $s$,  
we expect that $u$ and $d$, particularly $u$, 
quark fragmentation dominates the $\Lambda$ production 
in the high $p_\perp$ and large rapidity region.
However, $u$ or $d$ quark carries no spin of $\Lambda$ in the
SU(6) picture and only a small fraction of
the spin of $\Lambda$ in the DIS picture,
we expect that, just as in deeply inelastic scattering \cite{LXL2001}, 
the magnitudes of the $P_{\Lambda}$ obtained
in both models should be quite small, much smaller than those obtained 
in $e^+e^-$ annihilation near the $Z^0$ pole. 
The influence of heavier hyperon decay should be 
very important and the induced theoretical uncertainties in 
the calculations should be relatively large.

\begin{figure}[h]
\psfig{file=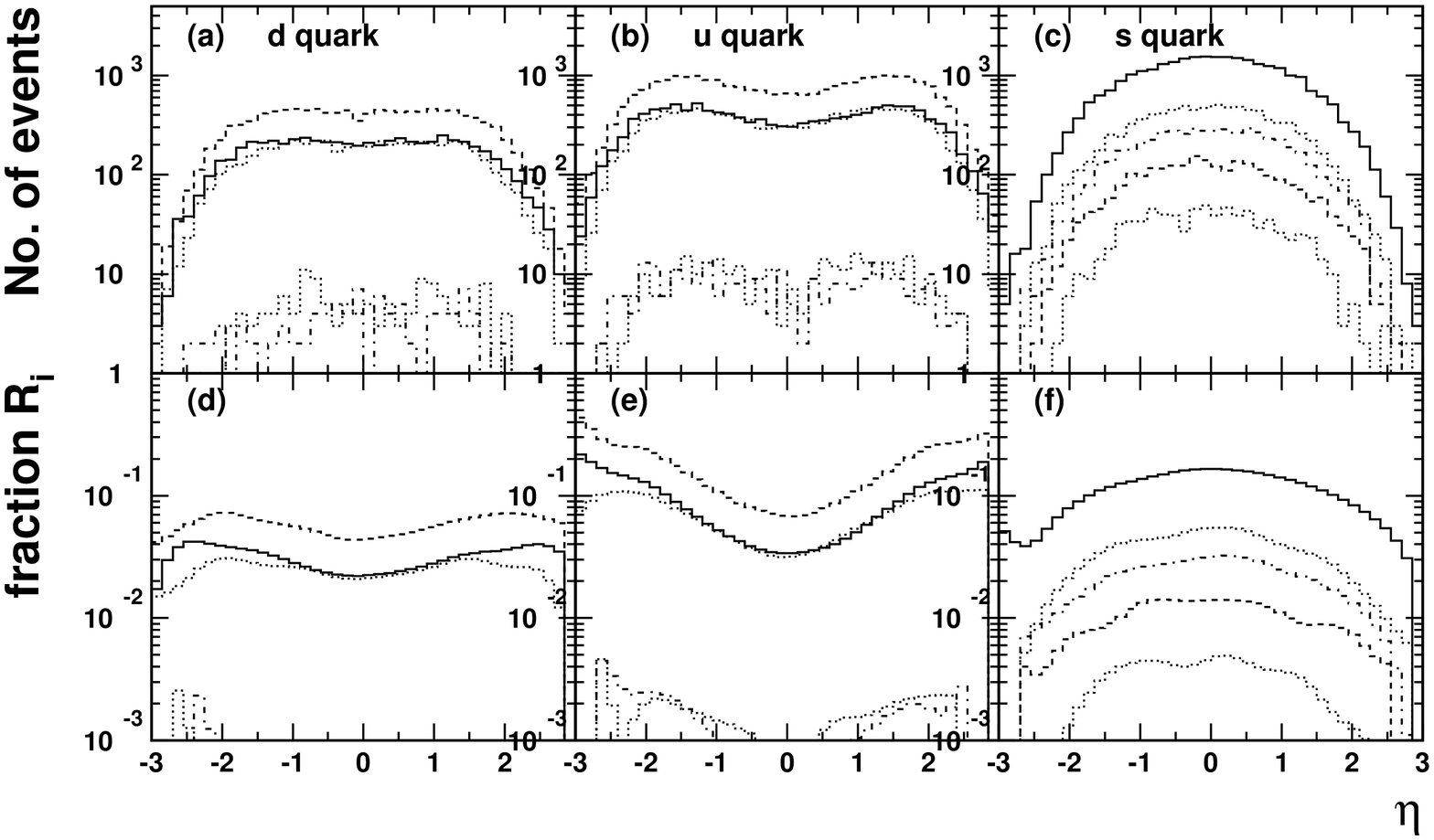,width=16cm}
\caption{Different contributions to $\Lambda$
production with $p_\perp$$>$13 GeV in events
originating from the $d$, $u$, or $s$ quark fragmentation
as a function of $\eta$ in $pp \to \Lambda X$
at $\sqrt s=500$ GeV.
In (a)-(c), we see the five types of contributions
from the fragmentation of $d$, $u$, or $s$ quark respectively.
The solid line denotes those which are directly produced
and contain the fragmenting quark;
the dashed, dash-dotted,
{\it upper} dotted, {\it lower} dotted,
denote those from the decay of
$\Sigma^0$, $\Xi$, $\Sigma^*$ and $\Xi^*$
which contain the fragmenting quark.
In (d)-(f), we show the fractions $R_i$
which are the ratios of the corresponding contributions
to the sums of all different contributions.}
\label{lamorg}
\end{figure}

\begin{figure}[ht]
\psfig{file=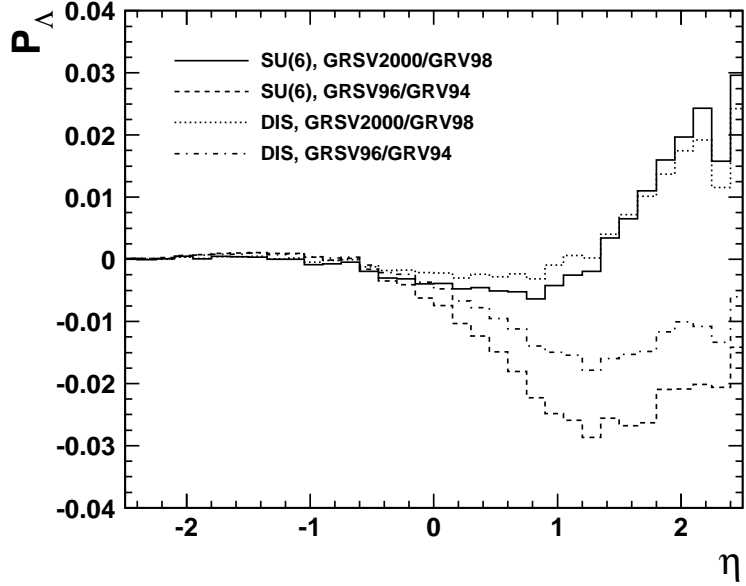,width=12cm}
\caption{Longitudinal $\Lambda$ polarization $P_{\Lambda}$
as a function of $\eta$ for $\Lambda$
with $p_\perp$$>$13 GeV
in $pp \to \Lambda X$at $\sqrt{s}=$ 500 GeV
with one of the beams is longitudinally polarized.
`GRSV2000/GRV98' denotes the Standard LO set of GRSV2000\cite{GRSV2000}
for polarized distribution functions and GRV98\cite{GRV98} LO set
for unpolarized distribution functions we used here;
`GRSV96/GRV94' denotes the Standard LO set of GRSV96\cite{GRSV96} and
GRV94\cite{GRV94} LO set correspondingly.}
\label{lampol}
\end{figure}

Using the event generator {\sc pythia},
we obtained the average numbers 
$\langle n^a_{\Lambda,f}\rangle$,
$\langle n^b_{\Lambda,H_j}\rangle$,
$\langle n^c_{\Lambda}\rangle$,
and $\langle n^d_{\Lambda}\rangle$
for $\Lambda$ production in $pp \to \Lambda X$.
By inserting them into Eq. (\ref{polh}), we obtain $P_\Lambda$ 
as a function of the pseudorapidity $\eta$.
We show the different contributions
to $\Lambda$ production at $\sqrt s$ =500 GeV
and $p_\perp$$>$13 GeV as functions of 
$\eta$ in Fig. \ref{lamorg} and 
$P_{\Lambda}$ in Fig. \ref{lampol}.
The moving direction of polarized beam proton is taken 
as the rapidity axis. 
From these results, we see in particular 
the following characteristics.

First, from Fig.\ref{lamorg}, we see that 
the contributions from $u$-quark fragmentation 
indeed dominate the $\Lambda$'s, 
in particular at large $\eta$. 
There are several direct consequences 
following from this : 
(i) the contributions from heavier hyperon decays are large, 
even larger than the directly produced contributions. 
To show this point more clearly, 
we present as an example in Fig. 9 the different contributions 
to $P_\Lambda$ in the SU(6) picture. 
We see that the magnitudes of the contributions 
to $P_\Lambda$ from $\Sigma^0$ and $\Sigma^{*0}$ decays 
are both much larger than those from directly produced 
$\Lambda$'s. 
We see that even the qualitative tendency are influenced 
very much by these decay contributions. 
(ii) the resulting $|P_\Lambda|$ is relatively small, 
it is less than 3\%. 
(iii) the differences between the results from SU(6) and those 
from the DIS picture are small. 

Second, the contributions from those hyperons which are 
directly produced and contain the fragmenting $q_f^0$'s 
[origin (a) mentioned in section 4.A]
or from the decays of the heavier hyperons which are directly 
produced and contain the fragmenting $q_f^0$'s 
[the type of heavier hyperons of origin (a)]
are much larger 
than the corresponding contributions in  the lepton-induced reactions. 
This is very interesting since these hyperons are all polarized 
and the polarizations are proportional to that of $q_f^0$. 
In fact, in the limiting case that there are only these 
two types of hyperons contribute, we have, 
\begin{equation}
P_{H_i}^{(lim)}=
{ {\sum\limits_f t^F_{H_i,f} P^{(q)}_f \langle n^a_{H_i,f}\rangle
+\sum\limits_{j} t^D_{H_i, H_j} P_{H_j}^{(a)} \langle n^b_{H_i, H_j}\rangle}
 \over
{\langle n^a_{H_i}\rangle +\langle n^b_{H_i}\rangle }}.
\label{polhlim}
\end{equation}
Here, we use the superscript $(a)$ in $P_{H_j}^{(a)}$ to denote 
the polarization of hyperon $H_j$ of only origin (a).
It is clear that, from Eq.(\ref{polh}), 
for such heavier hyperons, 
\begin{equation}
P_{H_j}^{(a)}=
{ {\sum\limits_f t^F_{H_j,f} P^{(q)}_f \langle n^a_{H_j,f}\rangle}
 \over
\langle n^a_{H_j}\rangle }.
\label{polhh}
\end{equation}
By inserting it into Eq.(\ref{polhlim}), we obtain that,
\begin{equation}
P_{H_i}^{(lim)}=
{ {\sum\limits_f [t^F_{H_i,f} \langle n^a_{H_i,f}\rangle
+\sum\limits_{j} t^D_{H_i, H_j}t^F_{H_j,f} 
\langle n^a_{H_j,f}\rangle R_{H_i, H_j} ]P_f^{(q) }
 \over
{\langle n^a_{H_i}\rangle +\sum\limits_{j}
\langle n^a_{H_j}\rangle R_{H_i, H_j} } }} ,
\label{polhlim2}
\end{equation}
where $R_{H_i,H_j}$ denotes the branching ratio for 
the decay $H_j\to H_i+X$.
We see clearly that, in this limiting case, 
the $P_{H_i}$ reflects 
more directly the polarization of the fragmenting quark $q_f^0$.
It is proportional to the $P_f^{(q)}$'s and the proportional constants 
depend on the picture for spin transfer and also the 
decay polarization transfer factors involved.
We see clearly that $P_\Lambda$ should depends rather 
strongly on $P_f^{(q)}$, which is determined by the 
spin transfer factors given in Table 1 and the 
polarized parton distribution functions. 
To see how sensitive it is, we used two different 
sets of parameterizations in the calculations. 
We see that the resulting difference in $P_\Lambda$ 
is indeed quite significant.

\subsection{Polarization of other hyperons}
      
From the results that we obtained in the last subsection,
we see that the contributions from the decay of heavier
hyperons to $\Lambda$ are large and 
the $\Lambda$ polarization is very small.
Since the contributions of different flavors to
the spins of the $J^P=(3/2)^+$ hyperons are
unknown yet in the DIS picture
and only models are known to calculate the spin transfer
in their decay processes,
there are considerably large theoretical
uncertainties in the results that we can obtain yet.
On the other hand,
as we have already seen\cite{LL2000,LXL2001} 
in $e^+e^-$$\to$$Z^0$$\to$$HX$ and in $\mu^-p\to\mu^-HX$,
the origins of other $J^P=(1/2)^+$ hyperons,
i.e., $\Sigma$ and $\Xi$,  are rather simple.
The decay contributions are much smaller than the corresponding
contributions for $\Lambda$ production.
The theoretical uncertainties in this case
can be significantly reduced.              
In particular in the large rapidity region, 
the $u$ quark fragmentation
plays the dominant role and $u$ carried most of the hyperon
$\Sigma ^+$'s spin.
Hence, the resulting $\Sigma^+$
polarization in $pp \to \Sigma ^+ X$ should 
be much larger than that for $\Lambda$.  
Furthermore, also because of the dominance of 
the $u$ quark fragmentation,
the production rate of $\Sigma^+$ should be comparable with
that of $\Lambda$, which implies that the statistics 
for studying $\Sigma^+$ should be similar to that of $\Lambda$.

Using the event generator {\sc pythia}, 
we calculate the different contributions to the inclusive
$\Sigma^+$ production in $pp$ collisions. 
Using Eq.(\ref{polh}), we obtain $P_{\Sigma^+}$ as a function of $\eta$ 
at $p_\perp$$>$13 GeV. 
The results are shown in Figs. \ref{sigorg} and \ref{hypol4} (a). 
Here, we see that the decay contribution from heavier 
hyperons is very small.
It takes only a few percents 
for $p_\perp$$>13$ GeV at $\sqrt{s}$=500 GeV.
We see also that the obtained 
$P_{\Sigma^+}$ is much larger than $P_\Lambda$
and the difference  between those results
based on the two different pictures are much more 
significant, in particular for large $\eta$.   
From these calculations, we also see again that the
contributions from the directly produced $\Sigma^+$'s which 
contain the fragmenting quark play the dominate role. 
In fact, they give more than 75\% 
of the whole produced $\Sigma^+$ at $p_\perp$$>$$13$ GeV. 
In particular for $\eta$$>$$1.5$, those directly produced 
and contain the fragmenting $u$-quarks take more than 93\%.
If we neglect all other contributions, we obtain 
the $P_{\Sigma^+}$ in this limiting case, i.e. 
the $P_{\Sigma^+}$ for those only from the origin (a) as, 
\begin{equation}
P_{\Sigma^+}^{(lim)}=P_{\Sigma^+}^{(a)}=
t^F_{\Sigma^+,u} P^{(q)}_u. 
\label{sigpollim}
\end{equation}
We see that it is directly proportional to the polarization 
of the $u$-quark, $P^{(q)}_u$, after the hard scattering 
and the proportional constant is just the
fragmentation spin transfer $t^F_{\Sigma^+,u}$. 
This provides indeed an excellent place to test different pictures 
for spin transfer in fragmentation processes. 
It may provide also 
a nice tool to study the polarization of the quarks.
Further studies along this line are under way\cite{Xu2002}.
From Fig.11, we see also that the results for 
$P_{\Sigma^+}$ 
from two different sets of parameterizations of 
the helicity parton distributions are rather small. 
This is because $P_{\Sigma^+}$ 
depends mainly on $\Delta u$, which is 
essentially the same in the two sets of parameterizations.

We also make similar calculations for other hyperons,
i.e., $\Sigma^-$, $\Xi^0$ and $\Xi^-$.
The results are shown in Fig. \ref{hypol4} (b)-(d).
We see that, $|P_{\Sigma^-}|$ exhibits similar behavior 
as $|P_{\Sigma^+}|$ but has opposite sign. 
This is because, 
similar to $\Sigma^+$ where $u$-quark fragmentation dominates,
$d$-quark fragmentation dominates $\Sigma^-$ production. 
But $\Delta d$ has opposite sign as $\Delta u$ in proton. 
Both $P_{\Xi^0}$ and $P_{\Xi^-}$ are determined mainly 
by the fragmentation of $s$-quark but with some influence 
from $u$ or $d$. 
Since $u$ or $d$ contributes negatively to the spin of $\Xi$, 
and $\Delta u$ is positive in proton but $\Delta d$ is negative, 
we should have a negative of $u$ quark fragmentation to $P_{\Xi^0}$ 
but a positive contribution from $d$ to $P_{\Xi^-}$. 
This makes $P_{\Xi^0}$ and $P_{\Xi^-}$ behave a bit different. 
The difference can be seen in Fig.11. 
All these different features can be used to make further tests 
to the SU(6) or DIS picture used for 
the spin transfer in fragmentation processes.

\section{Summary}

In summary, after a Monte Carlo 
analysis of the different contributions, 
we calculated the longitudinal polarizations for
different hyperons in high $p_\perp$ jets 
in $pp$ collisions with longitudinally polarized beams using 
two different pictures: the SU(6) and the DIS pictures.
The results show that the $\Lambda$ polarization
is rather small and has large contributions of 
different heavier hyperons decay.
On the contrary, the contribution to $\Sigma^+$ production
or that to $\Sigma^-$ production is rather pure
and the polarization is much larger.
Hence, the measurements of these polarizations 
can give a good test to the validity of the two pictures.
Such measurements can e.g. be carried out at 
RHIC in the near future.  
 
%%%%%%%%%%%%%%%%%%%%%%%%%%%%%%%%%%%%%%%%%%%%%%%%%%%%%%%%%%%%%%%%%%%%

\vspace{1cm}
\section*{Acknowledgments}

We would like to thank Li Shi-yuan and Xie Qu-bing 
and other members in the theoretical particle physics group of 
Shandong University for helpful discussions. 
This work was supported in part by
the National Science Foundation of China (NSFC) and 
the Education Ministry of China under Huo Ying Dong 
Foundation.

\newpage

\begin{table*}
\caption{Polarization transfer factors 
$D^{a\to c}_{(i)}(y)$ and
$D^{a\to d}_{(i)}(y)$ in the $i$-th kind of the 
elementary processes $ab \to cd$ 
when the incoming parton $a$ is longitudinally polarized. 
Here, $A$ and $B$ are defined as 
$A\equiv 3-10y+13y^2-6y^3+3y^4$ and $B\equiv 3-2y+y^2-2y^3+3y^4$. }
%\centering
\begin{tabular}{lll}
The elementary process &
\multicolumn{2}{c} {The polarization transfer factors\phantom{xxxxxxxxx} } \\ 
\cline{2-3}
$ab\to cd$ & $D_{(i)}^{a\to c}(y)$  &  $D_{(i)}^{a\to d}(y)$ \\ \hline
$(1)\ q_fq_f \to q_fq_f$ & 
$1-3y^4/A$ & $(2y-5y^2+6y^3)/A$  \\ \hline  
$(2)\ q_f {\bar q}_f \to q_f \bar{q}_f$ &
$1-3y^4/B$ & $ (2y-y^2+2y^3)$/B \\ \hline
$(3)\ q_fq_k({\rm or\ }{\bar q}_k) \to q_fq_k({\rm or\ }{\bar q}_k)$ 
&  1 &
$[1-(1-y)^2]/[1+(1-y)^2]$ \\ \hline
$(4)\ q_f {\bar q}_f \to q_k {\bar q}_k$ &
$[(1-y)^2-y^2]/[(1-y)^2+y^2]$ &
$[y^2-(1-y)^2]/[(1-y)^2+y^2]$  \\ \hline
$(5)\ q_f g \to q_f g $  &  1 & $[1-(1-y)^2]/[1+(1-y)^2]$ \\ \hline
$(6)\ g q_f \to g q_f $ &  1 & $[1-(1-y)^2]/[1+(1-y)^2]$ \\ 
\end{tabular}
\end{table*} 

\newpage

\begin{figure}[ht]
\psfig{file=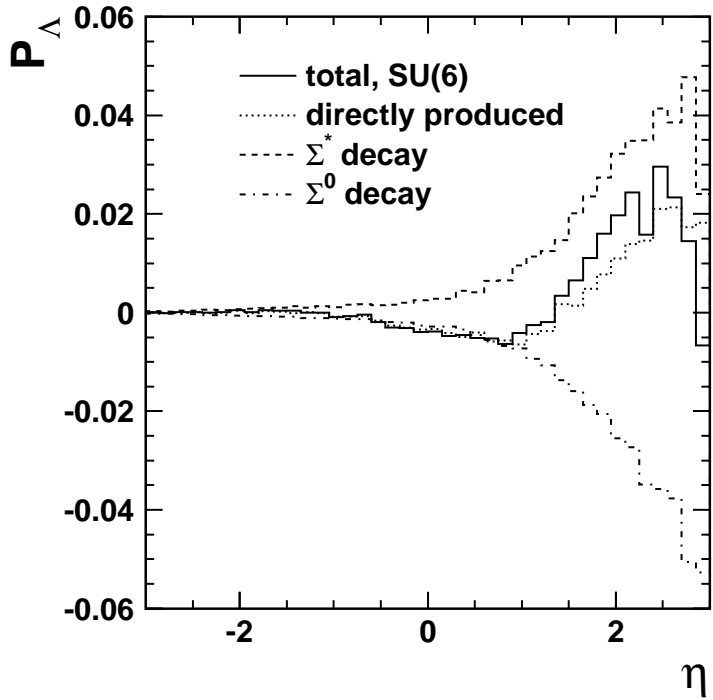,width=12cm}
\caption{Different contributions to $P_\Lambda$
in the SU(6) picture for $\Lambda$
with $p_\perp$$>$13 GeV
in $pp \to \Lambda X$ at $\sqrt{s}=$ 500 GeV 
in the case that the Standard LO set of GRSV2000\cite{GRSV2000}
for polarized distribution functions and GRV98\cite{GRV98} LO set
for unpolarized distribution functions are used.}
\label{lamdec}
\end{figure}

\begin{figure}[h]
\psfig{file=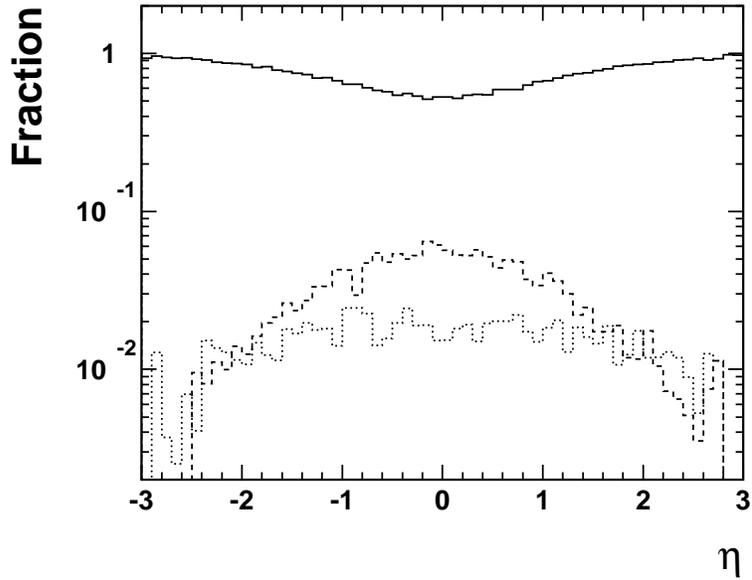,width=12cm}
\caption{Different contribution to $\Sigma^+$ production
with $p_\perp$$>$13 GeV
in $pp \to \Sigma^+ X$ at $\sqrt{s}=$ 500 GeV.
The solid and dashed lines are respectively the contributions
which are directly produced and contain the fragmenting quark
$u$ and $s$ respectively;
the dotted line corresponds to those which are originate from
the decay of polarized heavier hyperons.}
\label{sigorg}
\end{figure}       

\begin{figure}
\psfig{file=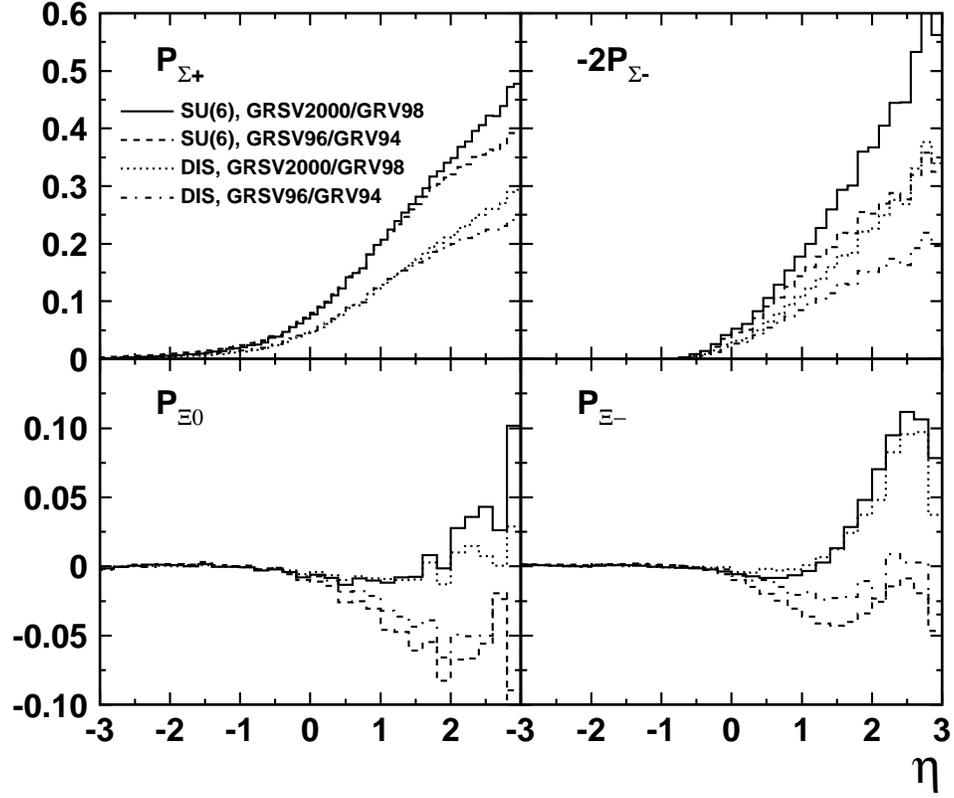,width=16cm} 
\caption{Polarizations for $\Sigma^+$, $\Sigma^-$,
$\Xi^0$ and $\Xi^-$
with $p_\perp$$>$13 GeV
in polarized $pp \to H X$ as a function of $\eta$
at $\sqrt{s}=$ 500 GeV.
Two sets of polarized distribution functions as same as for $\Lambda$
are used. }
\label{hypol4}
\end{figure}

\newpage

\noindent Table caption

\noindent
 Table 1: Polarization transfer factors 
$D^{a\to c}_{(i)}(y)$ and
$D^{a\to d}_{(i)}(y)$ in the $i$-th kind of the 
elementary processes $ab \to cd$ 
when the incoming parton $a$ is longitudinally polarized. 
Here, $A$ and $B$ are defined as 
$A\equiv 3-10y+13y^2-6y^3+3y^4$ and $B\equiv 3-2y+y^2-2y^3+3y^4$.

\vskip 2cm

\noindent Figure captions

\noindent
Fig.1: Contributions of the different parton level subprocesses
to $\Lambda$ production in $pp \to \Lambda X$.   

\vskip 0.3cm
\noindent
Fig.2: Contributions of the different parton level subprocesses
to $\Sigma^+$ production in $pp \to \Sigma^+ X$.      

\vskip 0.32cm
\noindent
Fig.3: The contributions of quark and gluon jets
to $\Lambda$ production in $pp\to\Lambda X$. 
 
\vskip 0.32cm
\noindent
Fig.4: The contributions of quark and gluon jets
to $\Sigma^+$ production in $pp\to\Sigma^+ X$.

\vskip 0.32cm
\noindent
Fig.5: The contributions of quark and gluon jets
to $\Lambda$ production (a) and
$\Sigma^+$ production (b) at $p_\perp$$>$3 GeV in $pp\to HX$.

\vskip 0.32cm
\noindent
Fig.6: 
Polarization transfer factors
$D^{a\to c}_{(i)}(y)$ and
$D^{a\to d}_{(i)}(y)$ in the $i$-th kind of the
elementary processes $ab \to cd$ in Table 1.

\vskip 0.32cm
\noindent
Fig.7:
Different contributions to $\Lambda$
production with $p_\perp$$>$13 GeV in events
originating from the $d$, $u$, or $s$ quark fragmentation
as a function of $\eta$ in $pp \to \Lambda X$
at $\sqrt s=500$ GeV.
In (a)-(c), we see the five types of contributions
from the fragmentation of $d$, $u$, or $s$ quark respectively.
The solid line denotes those which are directly produced
and contain the fragmenting quark;
the dashed, dash-dotted,
{\it upper} dotted, {\it lower} dotted,
denote those from the decay of
$\Sigma^0$, $\Xi$, $\Sigma^*$ and $\Xi^*$
which contain the fragmenting quark.
In (d)-(f), we show the fractions $R_i$
which are the ratios of the corresponding contributions
to the sums of all different contributions.

\vskip 0.32cm
\noindent
Fig.8:
Longitudinal $\Lambda$ polarization $P_{\Lambda}$
as a function of $\eta$ for $\Lambda$
with $p_\perp$$>$13 GeV
in $pp \to \Lambda X$at $\sqrt{s}=$ 500 GeV
with one of the beams is longitudinally polarized.
`GRSV2000/GRV98' denotes the Standard LO set of GRSV2000\cite{GRSV2000}
for polarized distribution functions and GRV98\cite{GRV98} LO set
for unpolarized distribution functions we used here;
`GRSV96/GRV94' denotes the Standard LO set of GRSV96\cite{GRSV96} and
GRV94\cite{GRV94} LO set correspondingly.  

\vskip 0.32cm
\noindent
Fig.9:
Different contributions to $P_\Lambda$
in the SU(6) picture for $\Lambda$
with $p_\perp$$>$13 GeV
in $pp \to \Lambda X$ at $\sqrt{s}=$ 500 GeV
in the case that the Standard LO set of GRSV2000\cite{GRSV2000}
for polarized distribution functions and GRV98\cite{GRV98} LO set
for unpolarized distribution functions are used.

\vskip 0.32cm
\noindent
Fig.10:
Different contribution to $\Sigma^+$ production
with $p_\perp$$>$13 GeV
in $pp \to \Sigma^+ X$ at $\sqrt{s}=$ 500 GeV.
The solid and dashed lines are respectively the contributions
which are directly produced and contain the fragmenting quark
$u$ and $s$ respectively;
the dotted line corresponds to those which are originate from
the decay of polarized heavier hyperons.

\vskip 0.32cm
\noindent
Fig.11:  
Polarizations for $\Sigma^+$, $\Sigma^-$,
$\Xi^0$ and $\Xi^-$
with $p_\perp$$>$13 GeV
in polarized $pp \to H X$ as a function of $\eta$
at $\sqrt{s}=$ 500 GeV.
Two sets of polarized distribution functions as same as for $\Lambda$
are used.

\end{document}